\documentclass[prl,twocolumn,showpacs,preprintnumbers,amsmath,amssymb]{revtex4}
\usepackage[latin1]{inputenc}
\usepackage[T1]{fontenc}

\usepackage{graphicx}
\usepackage{dcolumn}
\usepackage{bm}

\newcommand{\XeqY}[2]{$#1\!=\!#2$}

\begin{document}

\preprint{APS/N = 50 Shell gap}
\title{Evolution of the $N$=50 shell gap energy towards $^{78}$Ni}

\author{J.~Hakala}
\author{S.~Rahaman}
\author{V.-V.~Elomaa}
\author{T.~Eronen}
\author{U.~Hager} 
\altaffiliation[Present address: ]{TRIUMF, 4004 Wesbrook Mall, Vancouver, British Columbia, V6T 2A3, Canada}  
\author{A.~Jokinen}
\author{A.~Kankainen}
\author{I.~D.~Moore}
\author{H.~Penttil\"{a}}
\author{S.~Rinta-Antila} 
\altaffiliation[Present address: ]{Department of Physics, Oliver Lodge Laboratory, University of Liverpool, Liverpool L69 7ZE, UK} 
\author{J.~Rissanen}
\author{A.~Saastamoinen}
\author{T.~Sonoda} 
\altaffiliation[Present address: ]{Instituut voor Kern- en Stralingsfysica, Celestijnenlaan 200D, B-3001 Leuven, Belgium}
\author{C.~Weber} 
\author{J.~\"{A}yst\"{o}}\email{juha.aysto@phys.jyu.fi}
\affiliation{Department of Physics, P.O. Box 35 (YFL), FI-40014 
  University of Jyv\"{a}skyl\"{a}, Finland}

\date{\today}

\begin{abstract}

Atomic masses of the neutron-rich isotopes $^{76-80}$Zn,
$^{78-83}$Ga, $^{80-85}$Ge, $^{81-87}$As
and $^{84-89}$Se have been measured with high precision using
the Penning trap mass spectrometer JYFLTRAP at the IGISOL
facility. The masses of $^{82,83}$Ga, $^{83-85}$Ge,
$^{84-87}$As and $^{89}$Se were measured for the
first time. These new data represent a major improvement in the
knowledge of the masses in this neutron-rich region.  
Two-neutron separation energies provide evidence for the
reduction of the $N\!=\!50$ shell gap energy towards germanium
($Z\!=\!32$) and a subsequent increase at gallium ($Z\!=\!31$). The
data are compared with a number of theoretical models. An indication
of the persistent rigidity of the shell gap towards nickel
($Z\!=\!28$) is obtained.  
\end{abstract}

 	%% PACS, the Physics and Astronomy
			  		%% Classification Scheme.
\maketitle

Evolution of the shell structure of nuclei with extreme
neutron-richness is an increasingly important field of study in
nuclear structure physics and nuclear astrophysics. A specific issue
of major interest in recent years has been the question of whether
the magic nucleon numbers as known in the valley of stability survive
at large values of isospin \cite{do97, ot06} and the possible
role of the mutual support of magicities observed in or near the
valley of stability in the case of exotic nuclei \cite{ze83}.  
In this context, several experimental studies on excited states have
been performed to probe the evolution of the \XeqY{N}{50} neutron
shell far from 
stability. They have provided data on energies of the lowest excited
states of even-\emph{Z} \XeqY{N}{49} and \XeqY{N}{51} nuclei,
particle-hole excitations of even-even nuclei across the \XeqY{N}{50}
shell as well as B(E2) values for the ground state transitions of
$^{78,80,82}$Ge and $^{74,76,78,80}$Zn. The
experimental techniques employed include beta-delayed gamma-ray
spectroscopy of neutron-rich Cu, Zn and Ga 
\cite{pe06, ho81, wi07, ve07}, isomeric gamma-ray
spectroscopy of $^{76}\mathrm{Ni}$ \cite{ma05}, deep-inelastic studies
of \XeqY{N}{50} nuclei $^{87}\mathrm{Rb}$, $^{85}\mathrm{Br}$,
$^{84}$Se and $^{82}$Ge \cite{zh04} and of
excited fission fragments of $^{82}$Ge, $^{84}$Se,
$^{84,85}\mathrm{Br} $\cite{rz07, pr04, as06}. Very recently
experiments employing post-accelerated radioactive ion beams of 
$^{78,80,82}$Ge \cite{th05, pa05} and
$^{74,76,78,80}$Zn \cite{va07} have also been successfully
performed.

All current experimental results and particularly their interpretation
in the framework of the shell model indicate a weakening of the \XeqY{N}{50}
shell gap towards Ge (\XeqY{Z}{32}). Furthermore there is evidence
that the gap increases in the region of Zn \cite{va07}. The
interpretation, however, is very sensitive to the effective interaction
and core polarisation effects. Accordingly the quantitative
verification of the evolution of the shell gap energy has been missing
so far. Prior to the present work the experimental knowledge \cite{ame2003} of
two-neutron separation energies ($S_{2n}$) across \XeqY{N}{50}
indicated the reduction of the shell gap, i.e. the difference of
$S_{2n}$ values between the \XeqY{N}{50} and \XeqY{N}{52} isotones,
from \XeqY{Z}{40} down to \XeqY{Z}{34}. However, no conclusive trend
for the evolution of this gap further from stability can be determined
without new experimental results. 
The previous mass data \cite{ame2003} in this 
region are often associated with large systematic errors as observed
in our previous studies \cite{ha07a, ra07}.
In this Letter we present new data from high-precision mass
measurements of neutron-rich Zn, Ga, Ge, As and Se isotopes obtained
by employing a Penning trap technique \cite{bl06}. 

The measurements were performed using the JYFLTRAP Penning trap
setup \cite{jo06} which is connected to the Ion Guide Isotope
Separator On-Line (IGISOL) facility. The ions
of interest were produced in proton-induced fission reactions by
bombarding a 15 mg/cm$^{2}$ thick natural uranium target with
a proton or deuteron beam of 25 MeV energy. The fission reaction
products were stopped  in a helium-filled gas cell at a pressure of
about 200 mbar, and due to charge-exchange processes with the gas, the
ions ended up mostly singly charged. The helium flow then transported
the ions out of the gas cell towards the sextupole ion guide
(SPIG). 
After extraction the ions were accelerated to 30 keV and separated by
a $55^{\circ}$ bending magnet so that ions with the selected mass
number were guided into a gas-filled radio frequency quadrupole (RFQ)
cooler and buncher. The RFQ cooler was used to cool and
store the ions before releasing them in a short bunch towards the
Penning trap setup. 

\begin{figure}[htbp]
  \centering
  \includegraphics[width=0.48\textwidth]{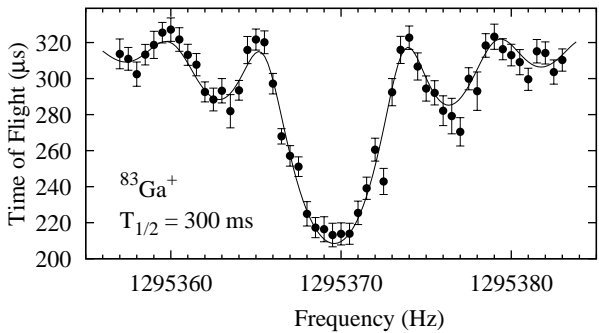}
  \caption{\label{fig:ga83} A typical time-of-flight (TOF) resonance
    from the JYFLTRAP setup for $^{83}\mathrm{Ga}^{\!+}$.}  
\end{figure}

In a Penning trap ions have three different eigenmotions: independent
axial motion and two coupled radial motions, magnetron $(\nu_{-})$ and
reduced cyclotron $(\nu_{+})$. The frequencies of the radial motions
sum to a true cyclotron frequency $\nu_c=\nu_{-} +  \nu_{+}$.
The JYFLTRAP setup consists of two cylindrical Penning
traps, the purification and precision trap, which are located inside
a 7 T superconducting magnet. 
In the purification trap filled with helium buffer gas, the
ions were further cooled by mass selective cooling and cleaned from
isobaric contaminants.
After being transported to the precision trap through a narrow
diaphragm, the magnetron radius of the ions was increased by a
dipole excitation. A quadrupole excitation 
was subsequently applied so that
the magnetron motion of the ions was converted to cyclotron motion,
whereby the ions gained radial energy. This energy gain was largest for
the resonance frequency $\nu_c\!=\!\frac{qB}{2\pi m}$ of the
ion cyclotron motion. Here \emph{q} and \emph{m} are the charge and
mass of the ions, and \emph{B} is the magnetic field.  
Finally, the ions were ejected towards a microchannel plate (MCP)
detector for measuring their time-of-flight (TOF). A typical TOF
resonance and the corresponding fit are shown in Fig.~\ref{fig:ga83}
where the fit is based on equations presented in \cite{konig}. 
The experimental setup and measurement technique are described in more
detail in \cite{ha07a, ra07}. 

Using the cyclotron frequencies of an unknown species ($m_{meas}$) and
a well-known reference ion ($m_{ref}$) and the mass of the electron
($m_{e}$), a precise mass determination for singly charged ions can be
made:
\begin{equation}
  m_{meas} = \frac{\nu_{c,ref}}{\nu_{c,meas}}
  \left(m_{ref} - m_{e}\right) + m_e.
\end{equation}

The masses of $^{76-80}$Zn, $^{78-83}$Ga,
$^{80-85}$Ge, $^{81-87}$As and $^{84-89}$Se
were measured during six different beam times (see Table \ref{tab:results}). 
Excitation times between 200 ms and 900 ms were used depending on the
half-life of each isotope. 
An excitation pattern with time-separated oscillatory fields (Ramsey
excitation) \cite{ge07} was used in the measurements in October
2007. 
Each isotope was measured typically three to five times and before and
after each measurement a similar measurement was performed for the
reference isotope $^{88}\mathrm{Rb}$ that has a mass excess of
$-82609.00\pm 0.16\ \mathrm{keV}$. 

The obtained average frequency ratios
$\bar{r}=\frac{\nu_{c,ref}}{\nu_{c,meas}}$ and the corresponding
mass-excess values are given in Table \ref{tab:results}. 
The systematic uncertainties due 
 to the presence of contaminating ions in the trap,
 the magnetic field fluctuations and
 the mass difference $A_{meas} - A_{ref}$ between the reference ion and
 the ion of interest
were taken into account in the cyclotron frequency ratio
determination. 
%%%
In order to reduce the effect due to ion-ion interactions in the trap
the count-rate was kept low by controlling the beam gate before the
RFQ cooler \cite{ha07a, ra07}. 
This effect was taken into account by applying the count rate class analysis
method described in \cite{ke03} when analysing the data.
In the case of low statistics this analysis was not possible and 
an estimate for the uncertainty was determined based on the analysis
made for the closest neighbouring isotope possible. 
The linear drift of the magnetic field was taken into account by the
interpolation of the reference cyclotron frequencies measured before
and after the ion of interest. The magnetic field fluctuation beyond
the linear drift was experimentally determined to be $3.2(2)\times
10^{-11}/\mathrm{min}$ 
\cite{ra07a} before the magnet quenched in June 2007 and 
$5.7(8)\times 10^{-11}/\mathrm{min}$ after re-energising the
magnet. This value, multiplied by the time difference between two
consecutive reference measurements, was added quadratically to each
frequency ratio uncertainty.
A mass-dependent systematic uncertainty of 
$7 \times 10^{-10}/\mathrm{u}$ was added quadratically to the final
frequency ratio uncertainty.  
For a consistency check we have compared our measurements of
$^{78}$Ga and $^{81}$As with previous high-precision
\cite{gu07, mo82} measurements. In addition we have measured
$^{79}$Ga in two different beamtimes and our values,
\hbox{-62547.8(22) keV} and \hbox{-62547.5(29) keV}, are in
excellent agreement with each other. 
In the case of $^{82}$As both the isomer and the ground state were
observed with their mass difference in good agreegment with
\cite{ga04}.

\begin{table}[htbp]
  \caption{\label{tab:results}Obtained cyclotron frequency ratios and
    resulting mass-excess values in keV from this work. 
    AME03 values based on extrapolation are marked with $\#$. 
    The index \emph{a} denotes the use of two-fringe Ramsey excitation.} 
  \begin{ruledtabular}
    \begin{tabular}{clll}
      Isotope & $\bar{r}=\frac{\nu_{c,ref}}{\nu_{c,meas}}$ & JYFLTRAP
      & AME03 \\[3pt] 
      \hline
      &&&\\[-6pt]
      $^{76}$Zn & 0.863745911(28) & -62303.8(23)$^{a}$ & -62140(80) \\
      $^{77}$Zn & 0.875163997(51) & -58789.5(42) & -58720(120) \\ 
      $^{78}$Zn & 0.886555123(33) & -57483.0(27)$^{a}$ & -57340(90) \\
      $^{79}$Zn & 0.897979777(33) & -53430.9(27)$^{a}$ & -53420(260$^{\#}$) \\
      $^{80}$Zn & 0.909376692(83) & -51650(7) & -51840(170) \\
      $^{78}$Ga & 0.886479142(37) & -63704.9(30)$^{a}$ & -63706.6(24) \\
      $^{79}$Ga & 0.897868446(23) & -62547.6(19)$^{a}$ & -62510(100) \\
      $^{80}$Ga & 0.909284208(35) & -59223.7(29) & -59140(120) \\
      $^{81}$Ga & 0.920678866(40) & -57628.0(33) & -57980(190) \\
      $^{82}$Ga & 0.932111398(29) & -52930.8(24) & -53100(300$^{\#}$) \\
      $^{83}$Ga & 0.943531430(32) & -49257.2(26) & -49390(300$^{\#}$) \\
      $^{80}$Ge & 0.909158285(25) & -69535.3(21) & -69515(28)  \\
      $^{81}$Ge & 0.920573067(25) & -66291.7(21) & -66300(120) \\
      $^{82}$Ge & 0.931958942(27) & -65415.1(22) & -65620(240) \\
      $^{83}$Ge & 0.943388317(30) & -60976.5(25) & -60900(200$^{\#}$) \\
      $^{84}$Ge & 0.954798023(39) & -58148.4(32) & -58250(300$^{\#}$) \\
      $^{85}$Ge & 0.966234558(45) & -53123.4(37) & -53070(400$^{\#}$) \\
      $^{81}$As & 0.920496845(37) & -72533.3(30) & -72533(6) \\
      $^{82}$As & 0.931901693(52) & -70103.1(43) & -70320(200) \\
      $^{82m}$As& 0.931903256(46) & -69975.1(38) & -70075(25) \\
      $^{83}$As & 0.943282162(34) & -69669.3(28) & -69880(220) \\
      $^{84}$As & 0.954703930(39) & -65853.5(32) & -66080(300$^{\#}$) \\
      $^{85}$As & 0.966111638(38) & -63189.1(31) & -63320(200$^{\#}$) \\
      $^{86}$As & 0.977538428(42) & -58962.1(34) & -59150(300$^{\#}$) \\
      $^{87}$As & 0.988954438(36) & -55617.9(30) & -55980(300$^{\#}$) \\
      $^{84}$Se & 0.954580662(24) & -75947.7(20) & -75952(15) \\
      $^{85}$Se & 0.965998990(32) & -72413.5(26) & -72428(30) \\
      $^{86}$Se & 0.977397491(31) & -70503.2(25) & -70541(16) \\
      $^{87}$Se & 0.988822450(27) & -66426.1(22) & -66580(40) \\
      $^{88}$Se & 1.000228663(41) & -63884.1(33) & -63880(50) \\
      $^{89}$Se & 1.011663571(46) & -58992.4(38) & -59200(300$^{\#}$)
    \end{tabular}
  \end{ruledtabular}
\end{table}

\begin{figure}[htbp]
  \centering
  \includegraphics[width=0.48\textwidth]{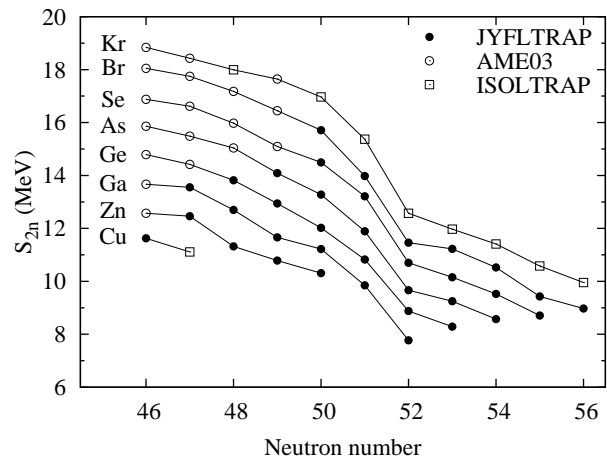}
  \caption{\label{fig:s2nplotn} Two-neutron separation energies for
    elements with proton number between \XeqY{Z}{29} and \XeqY{Z}{36}.}
\end{figure}

Figure \ref{fig:s2nplotn} displays the 
first-order derivative of the masses, the $S_{2n}$ values of
neutron-rich $Z\!=\,$29--36 isotopes calculated
from the mass-excess values measured by JYFLTRAP (Table
\ref{tab:results} and \cite{ra07, ra07a}) and ISOLTRAP
\cite{de07}. The new data from JYFLTRAP are shown as filled
circles. The occurrence of the closed neutron shell at \XeqY{N}{50} is
clearly visible from Kr (\XeqY{Z}{36}) all the way down to Ga
(\XeqY{Z}{31}), manifesting itself in a steep decrease of the $S_{2n}$
values for all nuclei with \XeqY{N}{51} and \XeqY{N}{52}. 

\begin{figure}[htbp]
  \centering
  \includegraphics[width=0.48\textwidth]{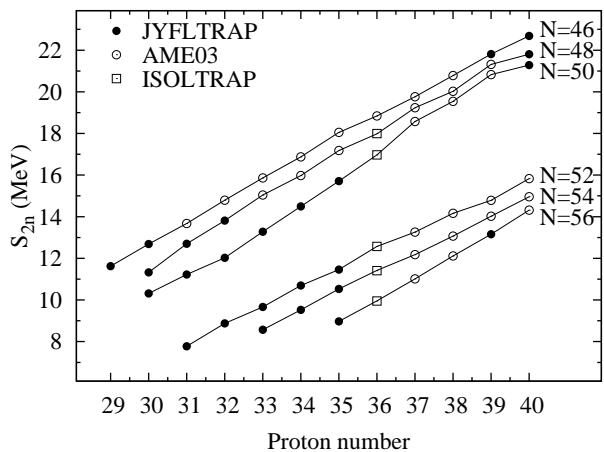}
  \caption{\label{fig:s2nplotz} Two-neutron separation energies for the
    region of interest plotted as a function of proton number.
    The evolution of the \XeqY{N}{50} shell gap is clearly visible.}
\end{figure}

In order to gain a more quantitative insight into the evolution of the
change of masses around shell closures, one can study the two-nucleon
gaps \cite{do97, be06} for neutrons or protons. For magic nuclei the
two-nucleon gap energy is approximated by twice the gap in the
single-particle spectrum providing a signature for the magicity
\cite{be06}. For this purpose, we have plotted in
Fig. \ref{fig:s2nplotz} $S_{2n}$ values for even \XeqY{N}{46}--56
isotones as a function of the proton number. The energy difference
between the \XeqY{N}{50} and 52 isotones, corresponding to a
two-neutron gap across \XeqY{N}{50}, reveals a trend for the lowering
of the shell gap energy towards Ge (\XeqY{Z}{32}). This also
corresponds to a minimum in the systematics of the first $2^{\!+}$
energies of known even-\emph{A} \XeqY{N}{50} isotones \cite{zh04,
  rz07, pr04, as06, th05, pa05, va07}, suggesting a maximal influence
of core polarisation effects at this proton
number. Fig. \ref{fig:s2nplotz} also indicates that the gap at
\XeqY{N}{50} opens up when moving further towards Ni (\XeqY{Z}{28})
implying its magic character.

Beyond the energy difference of \XeqY{N}{50} and 52 isotones, it is of
interest to notice that the otherwise smooth trend of $S_{2n}$ values
for each isotone is broken between the \XeqY{N}{48} and 50
isotones. Additional weakening of the two-neutron binding of the \XeqY{N}{50}
isotones between \XeqY{Z}{32} and 36 might arise, for example, from 
two-particle two-hole excitations across the shell gap. On the contrary,
when comparing the energy differences of the \XeqY{N}{46}, 48 and 52
isotones we observe a rather flat behaviour indicating a smoothness in
energies of the $\nu g_{9/2}$ and $\nu d_{5/2}$ (or $\nu s_{1/2}$)
neutron states across the entire range of the studied proton numbers.

\begin{figure}[htbp]
  \centering
  \includegraphics[width=0.48\textwidth]{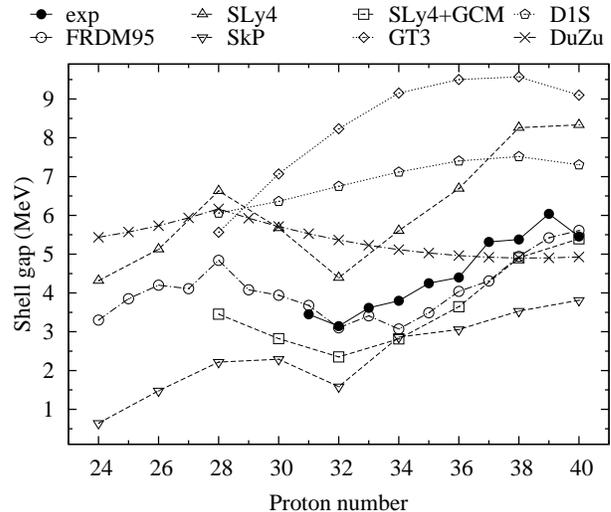}
  \caption{\label{fig:shellgap2} Evolution of the \XeqY{N}{50} shell
    gap and comparison to theoretical models.}
\end{figure}

The experimental two-neutron shell gap energies 
\hbox{$\Delta\!=\! S_{2n}(N\!=\!50) - S_{2n}(N\!=\!52)$} are compared
with selected theoretical models in Fig. \ref{fig:shellgap2}. 
The models chosen are the Finite Range
Liquid-Drop Model (FRDM) which is a well-known
sophisticated microscopic-macroscopic model \cite{mu95}, the empirical
model of Duflo and Zucker (DuZu) \cite{du02}, two spherical mean-field
calculations by Otsuka \textit{et al.} \cite{ot06,ot07} employing a GT3
tensor or a D1S interaction as well as three different
self-consistent mean 
field calculations in the framework of the density functional 
theory \cite{be06,st06,st03}. 
Both
mean-field calculations of Otsuka overpredict the gap by at least a
factor of two compared with the experimentally known range. The use of the
tensor-based interaction brings the gap clearly down to $^{78}$Ni. The
best agreement with the data is obtained by the FRDM
calculation. Finally, it is remarkable that the models based on
the density functional theory reproduce qualitatively the trend
observed in this experiment, a monotonic reduction from \XeqY{Z}{40}
and a minimum at \XeqY{Z}{32}. They also predict the increase in
the gap energy and therefore the strengthening of the magicity towards
$^{78}$Ni in compliance with the concept of mutual support of
magicities \cite{ze83}.

A closer look at Fig. \ref{fig:shellgap2} reveals that the
calculations of Stoitsov \textit{et al.} with SLy4 and SkP energy density
functionals \cite{st06, st03} agree in trend but differ in magnitude
rather strongly. This could, at least partly, be due to differences in
isoscalar effective mass in these two interactions which in turn
scales the single-particle level density in the vicinity of the Fermi
surface energy with the effective result of a larger gap with SLy4 and a
nearly vanishing gap for SkP at \XeqY{Z}{32}. On the other hand, the similar
calculation of Bender \textit{et al.} \cite{be06} employing the SLy4
interaction in the deformed basis and adding dynamical quadrupole
correlations brings the calculated values closer to those observed in
this experiment. 

In summary, we have measured the masses of approximately 30 neutron-rich
nuclei with unprecedented accuracy to probe experimentally the
magicity of the \XeqY{N}{50} neutron number far from stability. The data
indicates the persistence of this gap towards Ni (\XeqY{Z}{28}) with an
observed minimum at \XeqY{Z}{32}. This observation is in line with the
interpretation of recent spectroscopic data on low-lying excited
states of these nuclides in the framework of a shell model. Concerning the
binding energies, it is observed that the energy density functional
approach can reproduce qualitatively the observed trends in
two-neutron separation energies and shell gaps.
These high-precision experimental data provide a basis for improved
theoretical description of various separation energies of neutron-rich
nuclei needed, for example, in nuclear astrophysics. Moreover, the
future experiments will have to be pushed towards more exotic nuclei
such as $^{82}$Zn and $^{81}$Cu in the vicinity of $^{78}$Ni.

The authors wish to thank Professor Jacek Dobaczewski for inspiring
discussions.  

This work has been supported by the Academy of Finland under the
Finnish Centre of Excellence Programme 2000--2005 (Nuclear and
Condensed Matter Physics Programme at JYFL) and the Finnish Centre of
Excellence Programme 2006--2011 (Nuclear and Accelerator Based Physics
Programme at JYFL).

%%\bibliography{n50}

\end{document}